\documentclass{article}
\usepackage{blindtext}
\usepackage[a4paper, total={6in, 8in}]{geometry}

\usepackage{graphicx}% Include figure files
\usepackage[subrefformat=parens,labelformat=parens]{subfig}
\graphicspath{{./figures/}}
\usepackage{dcolumn}% Align table columns on decimal point
\usepackage{bm}% bold math
\usepackage[mathlines]{lineno}% Enable numbering of text and display math
\usepackage[export]{adjustbox}
\usepackage{color}
\usepackage{tabularx,array,booktabs}
\usepackage{caption}
\usepackage{mathtools}
\usepackage{amsmath,amssymb}
\usepackage{soul}
\usepackage{lipsum}
\usepackage{varioref}
\usepackage{algorithm,algpseudocode}
\usepackage{ulem}
\usepackage{siunitx}
\usepackage{comment}
\usepackage{upgreek}
\usepackage{url}

\usepackage{calc}
\usepackage{multirow,multicol}
\usepackage{pdfpages}
\usepackage[shortlabels]{enumitem}
\usepackage{rotating}
\usepackage{ulem}

\newcommand{\seba}[1]{\textcolor{blue}{\textit{Seba:}\ #1}}
\newcommand{\gia}[1]{\textcolor{red}{\textit{Gia:}\ #1}}
\newcommand{\marx}[1]{\textcolor{green}{\textit{Marx:}\ #1}}

\newcommand{\bb}[1]{\mathbf{#1}}
\newcommand{\bs}[1]{\boldsymbol{#1}}

\newcommand{\C}{\mathbb C}

\newcommand{\Comsol}{COMSOL Multiphysics\textsuperscript{\textregistered}}

\renewcommand{\seba}[1]{}
\renewcommand{\gia}[1]{}
\renewcommand{\marx}[1]{}

\renewcommand{\gg}[1]{}

\begin{document}

\title{High Bulk Modulus Pentamodes: the Three-Dimensional Metal Water}

\date{}

\author{
Giacomo Brambilla\thanks{Politecnico di Milano, Department of Mechanical Engineering, Milano, Italy, Corresponding author: giacomo.brambilla@polimi.it}
\and    Sebastiano Cominelli$^*$
\and    Marco Verbicaro$^*$
\and    Gabriele Cazzulani$^*$
\and    Francesco Braghin$^*$
}

\maketitle
\begin{abstract}

Despite significant advances in the field of phononic crystals, the development of acoustic metafluids that replicate the behaviour of liquids in three dimensions remains elusive. For instance, water -- the quintessential pentamode (PM) material -- has a bulk modulus two orders of magnitude higher than current state-of-the-art PMs.
% The unique requirements on elastic networks to be PM present considerable challenges.
The need for a low shear modulus inherently conflicts with the desire of high bulk modulus and density.

In this letter, we shed light on the limitations of existing PM geometries and propose an innovative shape for the links that constitute the network. Inspired by the kinematics of ropes, these links are constructed from thin fibres and demonstrate the potential to create PMs with properties akin to those of liquids.\\
As a prime example, we present the design of the first metamaterial that fully deserves the name \textit{3D metal water}, since its acoustic properties in the low frequency regime are indistinguishable with water.
Additionally, we highlight a shear band gap in the lattice dispersion diagram, and illustrate the influence of geometric parameters on the dynamic properties at higher frequencies.

This novel design of metafluids holds promise for applications requiring anisotropic materials such as acoustic lenses, waveguides, and cloaks.

\end{abstract}

\maketitle

\section{Introduction}
\label{sec:intro}

In the past two decades, significant progress has been made in passive control strategies for manipulating wave fields. Remarkable devices, such as superlenses~\cite{pendry2000negative}, cloaking~\cite{quadrelli2021experimental} and isospectral geometries~\cite{cominelli2023isospectral}, have been theorised and successfully implemented.
Their bizarre properties are accurately described by tailored mathematical models and attained by exploiting well-known phenomena that rely, for instance, on refraction, resonance and anisotropy~\cite{laude2020phononic}.

\noindent Beyond any doubt, the protagonists of such a revolution are the so-called metamaterials which are materials whose engineered microstructure mimics effective properties unavailable in nature.

Their design is based on diverse stratagems, but it mainly leverages space periodic structures whose dispersion relation is typical of crystals and that, for instance, exploits the well-known Bragg-scattering and locally resonance phenomena~\cite{laude2020phononic}.
A common method to study the quasi-static dynamics of a metamaterial relies on the so-called \textit{Low-Frequency Homogenisation}~(LFH)~\cite{antonakakis2014gratings}, an asymptotic method to compute the medium effective properties in the long-wave field approximation, based on the separation between micro and macro scales. Inherently, metamaterials are usually made of lattices whose cells are small compared to the scale of the device.

The two-scale design provides an unprecedented tool for engineers, that can now rely on palettes of materials with smoothly varying properties. However, even if several techniques are available to analyse a given metamaterial~\cite{antonakakis2014gratings}, few guidelines exist for the inverse design.\\
The design of sonic metamaterials presents a unique challenge due to the nature of acoustic waves, which propagate in media without a defined shape, such as gases and liquids. As a result, acoustic metamaterials often resort to 2D-like approximations, where solid inclusions are suspended in a liquid matrix~\cite{cominelli2022design,kadic2013metamaterials}, However, constructing devices based on such designs is often impractical, especially in 3D.
In principle, it is possible to use a solid device to transmit acoustic waves, as longitudinal pressure waves propagate in elastic media. However, these waves couple with transverse shear waves at the interfaces of the device, making their practical usage challenging unless specialised materials like pentamodes (PMs) are employed.

The concept of PMs was proposed for the first time in 1995 by Milton and Cherkaev~\cite{milton1995which}, who theorised a three-dimensional Face-Centered Cubic (FCC) Bravais lattice made by thin linkages of stiff material immersed in a soft matrix.
The effective elasticity tensor of these artificial crystals has five out of six null eigenvalues. Hence, these solids can oppose resistance to only one type of applied deformation. This is exactly what happens in fluids, where only hydrostatic stresses are sustained.
% In other words, these structures decouples longitudinal and transverse waves thanks to a bulk modulus infinitely larger than its shear moduli~\cite{kadic2012practicability}.
As a result, such materials are the perfect candidate to design \textit{metafluids}, being solid and, possibly, anisotropic.

\noindent PMs are usually designed considering their quasi-static dynamics and limiting to the frequency range where a full shear band-gap occurs~\cite{kadic2013anisotropic,li2019three,li2022elastic}; the former requirement in order to satisfy the hypothesis of LFH, the latter such that all the shear waves experience a vanishing propagation.

Acoustics is essential for communication and localisation in marine environments~\cite{zia2021state}, thus a PM capable of mimicking the acoustic behaviour of water represents an important achievement. This type of metafluid is known in literature as \textit{metal water}~\cite{norris2011metal}.
However, to the authors' knowledge, an elastic metamaterial with acoustic properties truly comparable to those of water has not yet been described.\\
Under the commonly made assumption of still, inviscid, and pure water at homogeneous temperature, the acoustic wave propagation is affected by the density $\rho_0$ and the bulk modulus $B_0$ of the liquid or equivalently by its sound speed $c_0=\sqrt{B_0/\rho_0}$ and acoustic impedance $Z_0=\sqrt{B_0\rho_0}$~\cite{ginsberg2018acoustics}.
Norris~\textit{et al}.\ \cite{norris2011metal} designed a two-dimensional PM (more properly called \textit{bimode}) featured by these properties and several experiments validated this result~\cite{zhao2016design,chen2015latticed,quadrelli2021experimental}.\\
3D pentamodes having sound speed similar to water exist~\cite{kadic2014pentamode,li2019three}, but their acoustic impedance is two order of magnitude lower than that of water, resulting in a different metafluid. Other PMs have been proposed based on the same cell arrangement~\cite{cushing2022design} or on different lattices~\cite{wei2023transformable,hu2023engineering}, but they all suffer from a low bulk modulus compared to water. In this work we investigate the reason behind it and propose a design of proper metal water.

An ideal PM -- featured by a null shear stiffness -- would flow away, hence some finite shear modulus improves the material stability. Nevertheless, attaining a high bulk modulus and low shear resistance is self-contradictory and leads to a challenging equilibrium.
Geometries based on double cone links are commonly adopted in literature since their practicability to achieve a negligible flexural-to-axial stiffness ratio~\cite{kadic2012practicability}. On this basis, the effect of the lattice parameters has been extensively studied~\cite{kadic2013anisotropic,li2019three,kadic2014pentamode,cushing2022design,huang2016pentamodal}.

In this letter, we delve into the theoretical prerequisites for lattice structures aspiring to achieve pentamode characteristics. Expanding upon the prevalent approach of employing double cones, commonly found in existing literature, we uncover a fundamental constraint on the effective bulk modulus of pentamode metamaterials based on such links.
% particularly when aiming for a specific shear-to-bulk stiffness ratio.
To address this limitation, we introduce an innovative shape for the lattice links, detailed in §\ref{sec:PM structure}, which effectively circumvents these constraints. Our proposed geometry demonstrates remarkable properties, boasting a shear stiffness approaching zero and a bulk modulus surpassing existing literature by two orders of magnitude.
Building upon this concept, we present the design of the inaugural 3D \textit{metal water} structure in §\ref{sec:unit cell}, focusing on its effective properties under quasi-static conditions and elucidating its dynamic behaviour through a comprehensive dispersion analysis. Finally, we draw conclusions and outline future avenues for exploration in §\ref{sec:conclusions}.

% However, the intriguing challenge of designing a PM which mimics water in all three spacial dimensions has never been faced. Questa frase non è del tutto vera

% This implies looking for a new arrangement for the unit cell of the PM such that both density $\rho_{0}$ and bulk modulus $B_{0}$ match the one of pure water.

% \seba{Altre cose che potremmo dire}

% Moreover, 3D manufacturability limits

% All in all, the attainability of the desired properties is often limited by the manufacturing processes at disposal.

\section{Limit of double-cones links}
\label{sec:PM structure}
% \input{Sections/2_PM_structure}

% \section{Links}
% \label{sec:links}

% Few precise rules exist for the design of PMs but a noteworthy effort has been spent by A.\ Norris \cite{norris2014mechanics} in outlining a simplified, yet detailed scheme for the analysis of PMs, interpreted as elastic networks.

\begin{figure} 
    \centering
    \subfloat[]
    {\includegraphics[height=.45\textwidth,trim={0 0 0 0}]{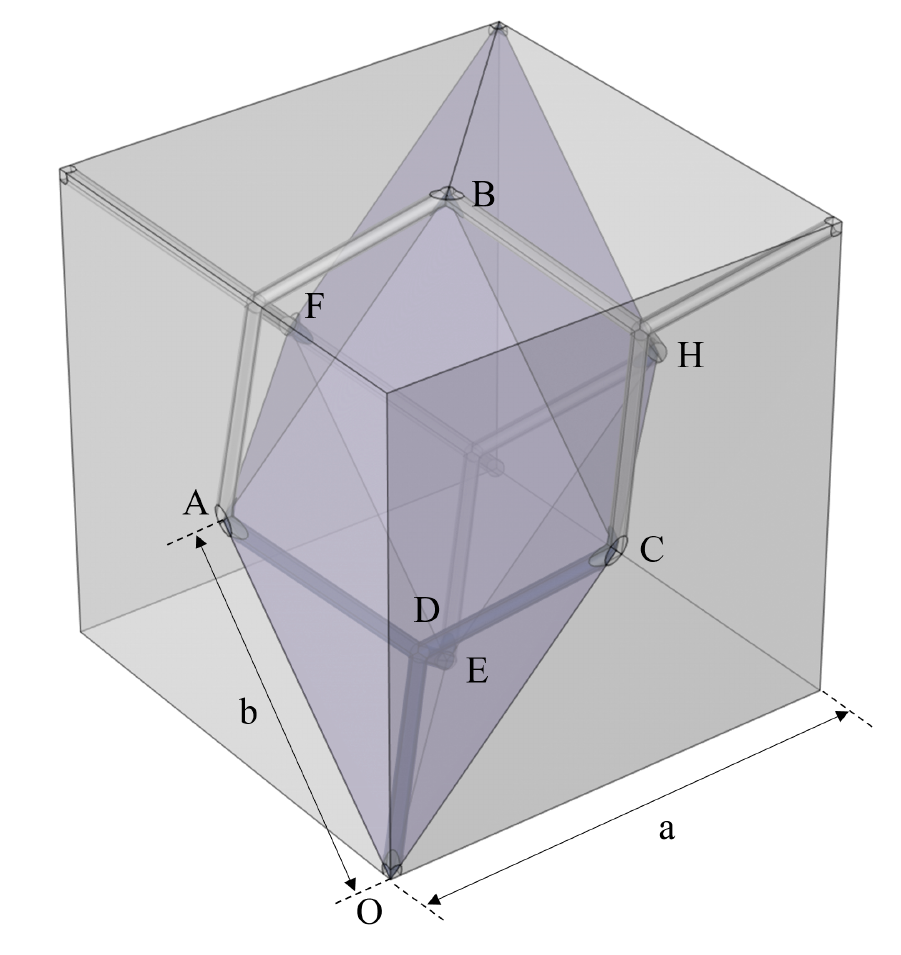}
    \label{fig:simplified_Cell_a}}
    \subfloat[]
    {\includegraphics[height=.45\textwidth,trim={0 0 0 0}]{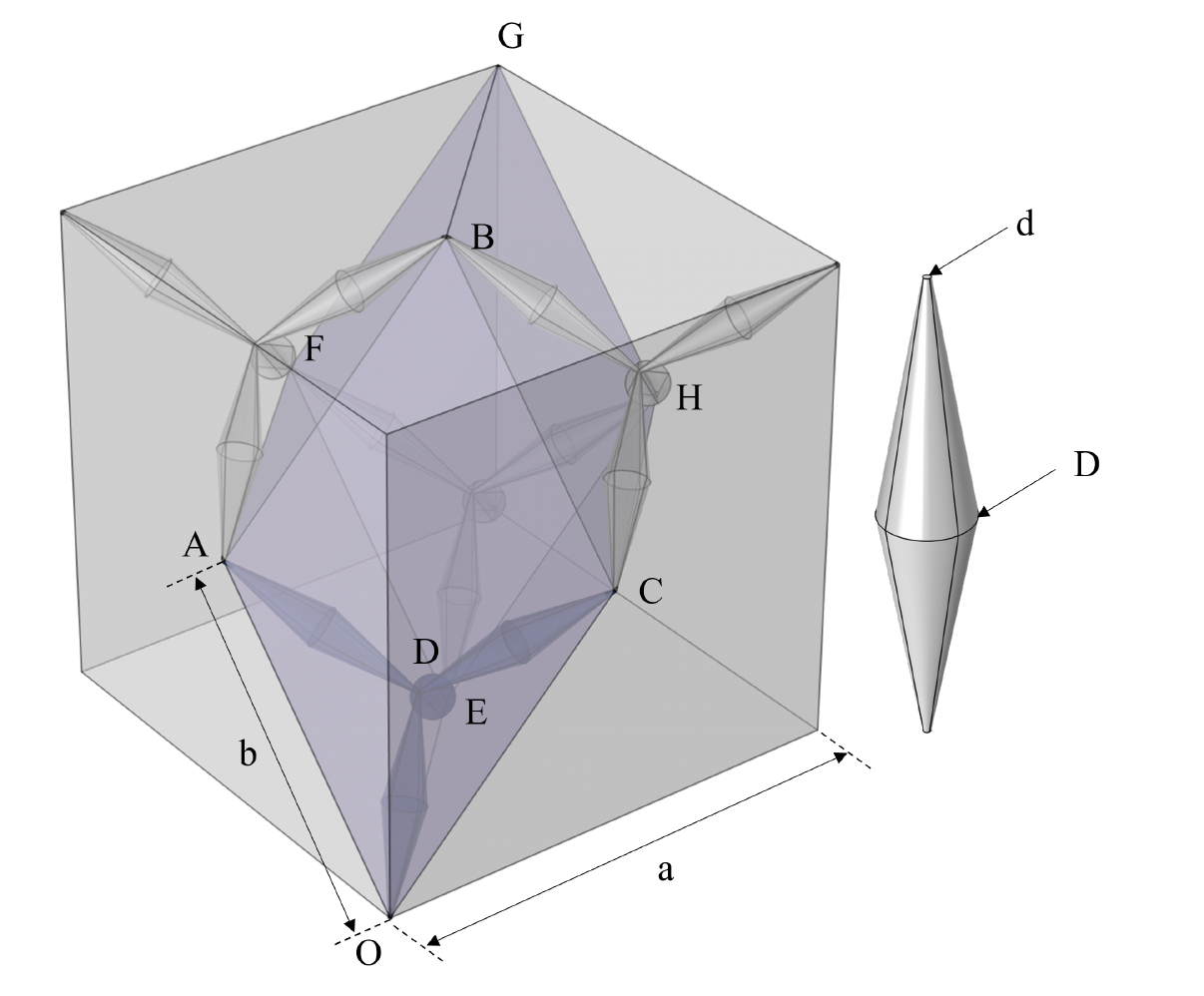}
    \label{fig:simplified_Cell_b}}
    \caption{Cubic unit cell containing four primitive unit cells. The trigonal trapezohedron highlighted in purple defines the elementary unit cell of the lattice. In (a) the lattice is composed by slender cylindrical links, while in (b) by double-cones.}
    \label{fig:simplified Cell}
\end{figure}

While the PM behaviour is easily appreciable from an intuitive standpoint, deep and rigorous mathematical investigations on the nature of such elastic lattices were proposed by many authors 
\cite{sun2012surface,deshpande2001effective,gurtner2014stiffest,norris2014mechanics}, by recasting the problem into the study of elastic networks made of links and joints.
A main result states that for an elastic network to be PM, each joint must connect exactly $Z=\gamma+1$ links whose deformation is dominated by stretching versus shearing (i.e., $K_{ax}/K_{sh}$ above a fixed amount, being $K_{ax}$ the axial stiffness and $K_{sh}$ the shear stiffness); where $Z$ is called coordination number and $\gamma\in\{2,3\}$ is the dimension of the space wherein the network extends.
The FCC lattice developed by Milton and Cherkaev is a good candidate to design a PM since its links are connected four by four, satisfying the constraint on the coordination number.
This geometry will be considered throughout this work for the sake of simplicity, but similar results apply to more complicated PM lattices proposed in the literature~\cite{wei2023transformable,hu2023engineering}. 
In particular, the geometry proposed by Li~\textit{et al.}~\cite{li2019three} uses the same concepts to build a PM whose lattice is hexagonal close-packed (HCP). Although this configuration comprises eight links instead of four, it shares many properties with the FCC arrangement.
A schematic drawing of the lattice is given in Figure~\ref{fig:simplified Cell}, where the fundamental directions of the lattice are defined by the edges $a$ of the cube wherein the elementary cell is outlined as a trigonal trapezohedron of edges $b$. Letters have been placed at the extremities of each linkage to facilitate the description.
% Moreover, three fundamental directions are defined by the edge of the cube together with the three face centres: indeed, they identify a trigonal trapezohedron that is nothing but the primitive unit cell of the considered FCC lattice.

Assuming all the links equal and their shear stiffness negligible, the effective bulk modulus is computed as~\cite{norris2014mechanics}:
\begin{equation}\label{eq:bulk cella}
    B_{eff} = \frac{4L^2K_{ax}}{9V},
\end{equation}
being $L$ the length of each rod and $V= 16\sqrt{3}/9\,L^3$ the volume of the unit cell.
When a specific bulk modulus is targeted, the problem reduces to find a proper geometry of the links, capable to attain the required axial stiffness $K_{ax}$ while minimising the shear stiffness $K_{sh}$.

In the first implementation by Kadic~\textit{et al.}~\cite{kadic2012practicability}, the links have the shape of two truncated and mirrored cones of minimum and maximum diameters $d$ and $D$, respectively. Refer to Figure~\ref{fig:simplified_Cell_b}.
% These double-cone elements are joint together in one extremity, while the others are located in correspondence of the vertices of a regular tetrahedron.
% Then, these elemental structures are properly oriented and repeated in space; 
In such a configuration, the connection between adjacent linkages takes place in correspondence of their small tips and pins are well approximated.
Thus the shear-to-bulk moduli ratio is contained
% ($<0.1$)
when $d$ is very small and the designed lattice maintains a PM-like behaviour at the expense of limited density and stiffness.
% , but the latter has the advantage of being more easily to handle when, e.g., the surface of a sphere has to be dressed \cite{li2019three}.

Throughout this section, we recast the commonly used double-cone shape into a family of solids of revolution and show that an intrinsic limit exists. Then a novel family of links is defined and modelled.

\subsection{Parametrized double cones}
\label{subsec:parametrizedDoubleCones}

% \marx{A mechanic network is considerable stretch-dominated when its links are characterised by a ratio $K_{ax}/K_{sh}$ above a fixed amount~\cite{norris2014mechanics}.}
% This, for instance, leads to study the stiffness of a truss by looking only at the axial forces within its links.
To have a stretch dominated network, the slender beams are a natural choice to build PMs and such a feature has been pushed to the limit in the literature using very thin links~\cite{kadic2012practicability,kadic2013anisotropic,li2019three,li2022elastic} and of different cross section~\cite{huang2016pentamodal}.
However, since the bulk modulus of a PM is proportional to the axial stiffness of its links, the slimmer the beam the smaller the PM bulk modulus. Then this strategy intrinsically limits the attainable bulk modulus of a PM to a value much lower than that of most liquids.

Under the hypothesis of a metallic truss (e.g., $E \approx \SI{100}{\giga\pascal}$) composed by cylindrical rods with cross-section diameter $d$, \eqref{eq:bulk cella} requires that the aspect ratio of each beam is $d/L \approx 0.44$ when the bulk modulus of water ($B_0=\qty{2.2}{\giga\pascal}$) is targeted, which cannot be considered slender.
On top of that, when the geometry of the lattice is considered, the high interpenetration occurring in correspondence of each node is no longer negligible, thus the equivalent length of the links reduces and the approximation of stretch-dominated deformations breaks. As a result, the lattice is no longer a PM.\\
Such an impasse is slightly improved by having non-uniform rods, for example if we take into account a solid of revolution whose diameter $d = d(\chi)$ varies along the axis coordinate $\chi\in[0,1]$.
In this more generic case, under the assumption of an Euler-Bernoulli slender beam, the following holds:
\begin{subequations}
\label{eq:DefinitionStiffness}
\begin{equation}
    \label{eq:DefinitionStiffnessAx}
    K_{ax}^{-1} = \int_0^1\frac{L}{EA(\chi)} \, d\chi = \int_0^1\frac{4L}{E\pi d^2(\chi)} \, d\chi,
\end{equation}
\begin{equation}
    \label{eq:DefinitionStiffnessSh}
    K_{sh}^{-1} = \int_0^1\frac{L^3\chi^2}{EI(\chi)} \, d\chi = \int_0^1\frac{64 L^3\chi^2}{E\pi d^4(\chi)} \, d\chi,
\end{equation}
\end{subequations}
where $E$ and $I(\chi)$ are the Young's modulus and the second moment of inertia of the beam.
In the ideal case, we would like to obtain $K_{sh}\to0$. This is possible only if the integrand function in \eqref{eq:DefinitionStiffnessSh} is unbounded at least in one point, that is, if exists $\chi^*\in[0,1] \colon d(\chi)/\chi^2\to 0$, for $\chi\to \chi^*$.
So, at the leading order, the diameter is described by $d(\chi) \approx \mu |\chi-\chi^*|^\alpha$, for $\chi$ about $\chi^*$, $\alpha>0$.
Let us study the convergence of the two integrals in \eqref{eq:DefinitionStiffness} as a function of the parameter $\alpha$.

\noindent We stick to reality by considering a symmetric configuration where the diameter goes to zero in correspondence with the tips, i.e.:
\begin{equation*}
    \label{eq:DiameterParabolae}
    d(\chi) = 
    \begin{dcases}
    \mu \chi^\alpha, \qquad 0\leq \chi\leq \frac{1}{2}\\
    \mu (1-\chi)^\alpha, \qquad \frac{1}{2} < \chi\leq 1
    \end{dcases},
\end{equation*}
where the parabolic shape describing the proximity of $\chi^*$ has been extended through the entire beam.
Note that the commonly used double-cone shape is recovered for $\alpha = 1$.\\
Thanks to the symmetry about $\chi=1/2$, the axial stiffness is computed as:
\begin{equation*}
    K_{ax}^{-1}\,\,\frac{E\pi\mu ^2}{8L} = \lim_{z\to 0^+} \int_z^{1/2} \frac{1}{\chi^{2\alpha}}\, d\chi.
\end{equation*}
% where $\hat K_{ax}\coloneqq $ is the non-dimensional axial stiffness.
A non-null value is desired for $K_{ax}$, so $\alpha<1/2$.
The shear compliance is computed as:
\begin{multline}
    \label{eq:convShear}
    K_{sh}^{-1}\,\,\frac{E\pi\mu ^4}{64L^3} = \lim_{z\to 0^+} \int_z^{1/2} \frac{\chi^2}{\chi^{4\alpha}} d\chi
    + \lim_{w\to 1^-} \int_{1/2}^w \frac{\chi^2}{(1-\chi)^{4\alpha}} d\chi \\=
    \lim_{z\to 0^+} \int_z^{1/2} \frac{2\chi^2}{\chi^{4\alpha}} - \frac{2\chi}{\chi^{4\alpha}} + \frac{1}{\chi^{4\alpha}}\, d\chi.
\end{multline}
$K_{sh}$ is null if at least one of the three integrals diverges, that is, $\alpha>1/4$.
Thus, any value of $\alpha \in [1/4,\, 1/2)$ is a good candidate for composing a PM lattice.

\begin{figure}
    \centering
    \subfloat[]{\includegraphics[height=0.4\textwidth,trim= 20 0 20 0,clip]{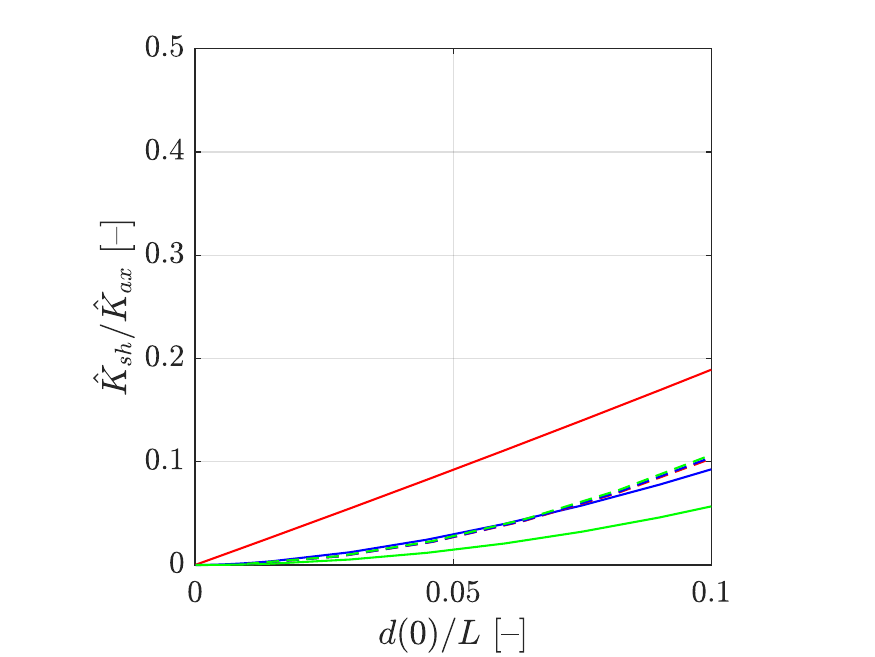}}
    \label{fig:k_sh_k_ax_a}
    \subfloat[]
    {\includegraphics[height=0.4\textwidth,trim=20 0 20 0,clip]{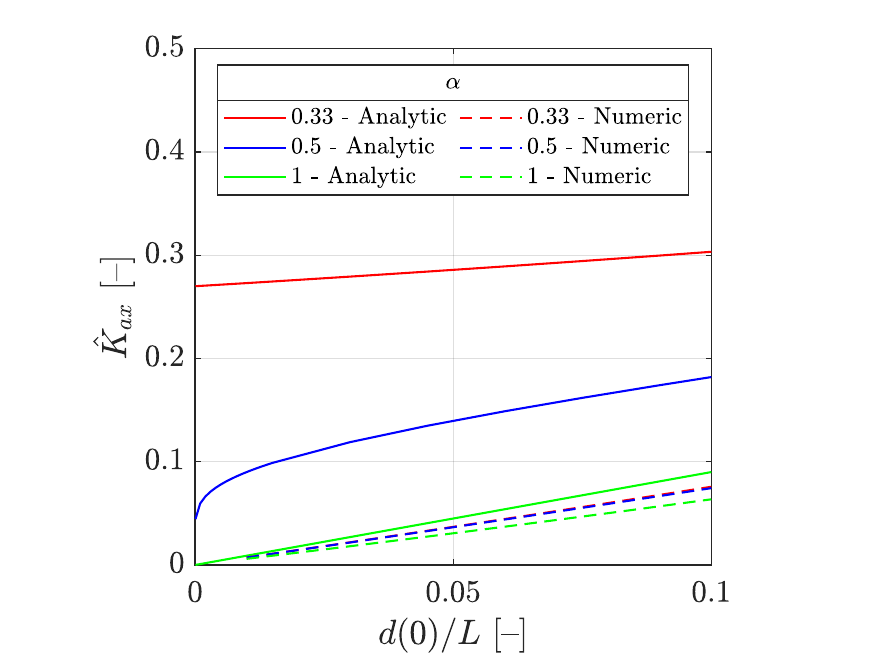}}
    \label{fig:k_sh_k_ax_b}
    \caption{Comparison between the analytical and numerical estimation  of the parabolic beam in terms of (a) $\hat K_{sh}/\hat K_{ax}$ and (b) $\hat K_{ax}$. In both cases a fix value $d(\chi/2)=0.9\,L$ has been chosen to show what happens when a high bulk modulus is pursued.}
    \label{fig:stiffness}
\end{figure}

In practice, a beam must have a positive diameter so the shear stiffness cannot be null.
The computations considering a finite diameter are available in the supplementary material, where $\hat K_{ax}$ and $\hat K_{sh}$ are defined in equation (SM.4).
and lead to the curves shown in Figure~\ref{fig:stiffness}.\\
In theory, given a maximum threshold on the ratio $K_{sh}/K_{ax}$, $\alpha<1$ permits higher $K_{ax}$.
However, the comparison with a numerical solution based on finite elements (dashed lines) shows that the improvement is very limited. This important difference is due to the Euler-Bernoulli approximation that no longer holds at the extremities of the parabolic beam, where the cross-section rapidly varies along the beam axis.
% Thus, the hypothesis of stretch-dominated deformations is addressed by setting a maximum threshold on the ratio $K_{sh}/K_{ax}$~\cite{norris2014mechanics}.

\subsection{Sheaf of beams}
Thereby it is evident that the adoption of double-cones links is of little interest when the bulk modulus of a liquid is targeted.
The double-cones are commonly adopted to tune the bulk and density: shear-compliant links are achieved thanks to their thin extremities while the mass is obtained by increasing the thickness at the centre of the links.\\
In such a configuration, the stretch-dominated condition is enhanced by making the nodes of the lattice similar to hinges, such that only axial forces are exchanged by the links. This concept works to mimic water in the 2D approximation, but we just proved to be limited in 3D because it is a bottleneck for the axial stiffness.\\
To overcome such a limitation, we propose a new design paradigm of the PM. We note that the effective stress tensor $\bs\sigma_h$ of a mechanic network is computed as $\bs\sigma_h=\sum_i \bb r_i\otimes\bb f_i$ \cite{norris2014mechanics}. In particular, the torques occurring at the nodes have no effect, as we precisely prove in the supplementary material.
Hence we relax the hinge-hinge kinematics designing a link acting as a slider instead.
Inspired by the kinematics of ropes, we propose a layout where links are composed by several parallel fibres. 
% \gccanc{a family of links composed by several fibres instead.}
The outcome is a mechanical element whose shear stiffness decreases by increasing the number of fibres.
% depends on the whole length of the link, both when shear or stretch actions are applied, 

Cables and ropes rely on woven and twisted sheaves of small fibres that tighten together when the cable is stretched and slide one on another when it is bent \cite{backer1952mechanics}.
Such a mechanism shows remarkably low-shear and high-axial stiffness, but it constitutes a challenge for the production at small scales.\\
In the first place, the structure of a cable requires to produce each fibre apart and bundle them together in two subsequent stages, such that the sheaf is tight;
this becomes really challenging when a significant number of links is desired, as in a lattice.
This issue is overcome thanks to additive manufacturing processes, with the drawback that some clearance must be considered in between two fibres to avoid them to stick one another during their production.

\begin{figure}
    \centering
    \subfloat[]{\includegraphics[width=0.45\textwidth]{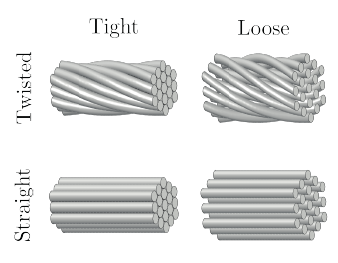}
    \label{fig:sheaf-a}}
    \quad
    \subfloat[]{\includegraphics[width=0.45\textwidth,trim=0 -10 0 0]{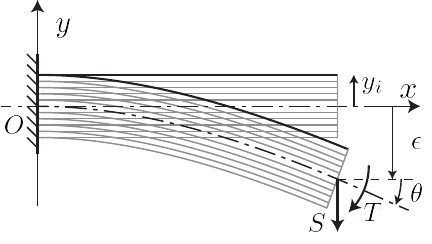}\label{fig:sheaf-b}}
    \caption{(a) composition of a sheaf of fibres and (b) schematic of its deformation. }
    \label{fig:sheaf}
\end{figure}

\noindent In the second place, the high shear compliance of a cable takes advantage of the fibres' twisting: when a cable undergoes bending, each fibre sustains simultaneously elongation and compression, that cancel out as a result \cite{backer1952mechanics}.
However, when a twisted bundle is loose, the axial stiffness of the cable drastically decreases because the twisted shape of each fibre is easily stretched or compressed, look at Figure~\ref{fig:sheaf} for a schematic drawing.

Hereby, we study the elastic response of a link composed by many thin beams parallel one another.
Such a collection of beams will be equivalently called sheaf/bundle of beams/fibres. For the sake of simplicity we limit here the analysis to the 2D deformation depicted in Figure~\ref{fig:sheaf-b}, that is sufficient to comprehend the principle of such a mechanism. The full 3D study is reported in the supplementary material. \\
Let $N$ be the number of fibres composing a sheaf, each one being a cylinder with height $L$ and cross-section diameter $d\ll L$, such that it is considered as a slender beam. The cross section area of each fibre $i$ is $A = \pi d^2/4$, then from~\eqref{eq:DefinitionStiffness} we have $K_{ax}^i= \frac{EA}{L}$ and $K_{sh}^i=\frac{3EA^2}{L^3}$.
Let us consider a 2D Cartesian reference system placed on the cross-section plane and centred on the neutral axis of the sheaf,
% barycentre of the cross-section area,
where $y_i$ is the vertical coordinate of the centre of the $i$\textsuperscript{th} beam.
% i.e.\ $\sum_{i=1}^N \bb y_i = \bb 0$.\\
We assume that all the fibres are rigidly connected in their two extremities; this approximates two rigid terminations of the sheaf.
We derive the kinematic model applying the \textit{direct stiffness method} \cite{felippa2001historical} such that the compatibility of the beams clamped on the two planes is easily imposed.

Without loss of generality, we limit ourselves to study the deformation of the sheaf in a plane containing its neutral axis.
With reference to Figure~\ref{fig:sheaf-b}, let the left end of the sheaf be grounded while the transversal displacement $\epsilon$ and the rotation $\theta$ of the right end are controlled. On this extremity, we measure the required transversal force $S$ and torque $T$ needed to impose a deformation.
The direct stiffness method, applying to linear systems, allows the superposition principle thus we independently vary $\epsilon$ and $\theta$ and measure the forces and the torques required to deform the $i$\textsuperscript{th} fibre.\\
% for each and every beam, then we sum all the forces and the torques required to deform the beams
% to measure the stiffness of the sheaf.\\
If $(\epsilon,\theta)=(1,0)$, the fibre is subject to shear $S_i^\epsilon=\frac{12E}{L^3}I$ and torque $T_i^\epsilon=\frac{6E}{L^2}I$, where $I$ is the second moment of area of the fibre cross section.
Conversely, when $(\epsilon,\theta)=(0,1)$, $S_i^\theta=\frac{6E}{L^2}I$ and  $T_i^\theta=\frac{E}{L}(4I + Ay_i^2)$. Hence, the stiffness matrix of the sheaf is obtained summing the forces of each fibre in case of a general deformation $(\epsilon,\theta)$:
\begin{equation}
    \begin{bmatrix}
    S \\ T
    \end{bmatrix}
    =\sum_{i=1}^N
    \begin{bmatrix}
        S_i^\epsilon & S_i^\theta
        \\[4pt]
        T_i^\epsilon & T_i^\theta
    \end{bmatrix}
    % \begin{bmatrix}
    %     \frac{12E}{L^3}I & \frac{6E}{L^2}I
    %     \\[4pt]
    %     \frac{6E}{L^2}I & \frac{E}{L}(4I + A y_i^2)
    % \end{bmatrix}
    \begin{bmatrix}
        \epsilon \\ \theta
    \end{bmatrix}
    = 
    \begin{bmatrix}
        N\frac{3E}{\pi L^3}A^2 & N\frac{3E}{2\pi L^2}A^2
        \\[4pt]
        N\frac{3E}{2\pi L^2}A^2 & \frac{E}{L}(N\frac{1}{\pi}A^2 + A\sum_i^N y_i^2)
    \end{bmatrix}
    \begin{bmatrix}
        \epsilon \\ \theta
    \end{bmatrix},
\end{equation}
where the equivalence $I = \frac{1}{4\pi} A^2$ holding for a circle has been used. Hence, given the beams distribution $\{y_i\}_1^N$, the stiffness matrix is computed.

The axial stiffness of the sheaf is of major importance for the effective bulk modulus of the lattice, and it is easily computed as $K_{ax}=\frac{EA_{tot}}{L}$, where $A_{tot}=nA$ is the total cross-section area.
Let us compute $\bb K$ when a specific axial stiffness $\tilde K_{ax}$ is targeted and the number of fibres $N$ is arbitrarily large.
Given $L$ and $E$, the total area is $A_{tot}=\frac{L\tilde K_{ax}}{E}$ and the area of each fibre is computed as $A=A_{tot}/N$. By substitution, the stiffness matrix becomes
\begin{equation}
    \bb K = 
    \begin{bmatrix}
        \frac{3E}{\pi L^3 N}A_{tot}^2 & \frac{3E}{2\pi L^2N}A_{tot}^2
        \\[4pt]
        \frac{3E}{2\pi L^2 N}A_{tot}^2 & \frac{E}{L}(\frac{1}{\pi N}A_{tot}^2 + \frac{A_{tot}}{N}\sum_i^N y_i^2)
    \end{bmatrix}
    \underset{N\to\infty}{=}
    \begin{bmatrix}
        0 & 0 
        \\[4pt]
        0 & \frac{E}{L}I_\infty
    \end{bmatrix},
\end{equation}
because three terms out of four are proportional to $N^{-1}$, independently of the fibres distribution; conversely, the fourth term tends to the lower limit $\frac{E}{L}I_\infty$, where $I_\infty\coloneqq\lim_{N\to\infty} \frac{A_{tot}}{N}\sum_i^N y_i^2$ is the only term that depends on the spatial distribution of the fibres.

\noindent On top of that, for a large $N$, the hypotheses on the slenderness of each fibre is fully verified. Obviously, some practical constraints exist, for instance a hair-like fibre cannot support any relevant axial compression since buckling is critical for thin beams.
Such a compromise is even tighter when the manufacturing limitations are considered. In this respect, the freedom about the distribution of the fibres plays an important role.

Since for large $N$ the only non-null term is $T_i^\theta$, a torque occurring at one extremity of the sheaf is balanced only by the torque occurring on the other extremity and no shear force arises. Then the PM behaviour is preserved since the quasi-static effective stress of the lattice is influenced only by the forces transmitted node by node, and not by the torques.
% ~\cite{norris2014mechanics}
The reader is referred to the supplementary material for a proof.
Note that a three-dimensional sheaf exploits a torsional stiffness that vanishes proportionally to $N^{-1}$, as shown in the supplementary material. The same conclusion holds for the 3D lattice. 
The role of transmission of torque affects the high-frequency regime, as shown in the dispersion diagram in the next section.

Thanks to this new link, we remove the theoretical upper limit on the attainable bulk-to-shear moduli ratio of a PM; the constraint is now given by geometrical and manufacturing issues. In the following section, we tackle such problems by designing the so called \textit{metal water}.

\section{Metal Water unit cell}
\label{sec:unit cell}

% \subsection{Unit cell geometry}
% \label{subsec:unitCellGeom}

With reference to Figure~\ref{fig:simplified Cell}, the four links $OD$, $AD$, $CD$, and $DE$ are now comprised of bundles of thin, parallel, cylindrical beams closely spaced together. Consequently, the construction complexity primarily stems from modelling each sheaf and devising an efficient method for connecting them.\\
A precise axial stiffness $K_{ax}$ is required to each sheaf to obtain a PM whose bulk modulus approximates the one of water. It depends on the sheaf cross-section area $A_{tot}$ that, in turn, defines the number of its fibres, once their minimum size is chosen.
The fibres of each bundle are arranged according to a hexagonal pattern
% , as shown in Figure \ref{fig:bundleCrossSection}, 
to enhance the compactness.
Note that
% , due to symmetries in the primitive unit cell, the lattice is \seba{orthotropic ??} and
the four linkages are identical, thus the same construction mechanism is replicated four times. For more details the reader is referred to the supplementary material.

\begin{figure} 
    \centering
    \subfloat[][]{
    \includegraphics[height=.5\textwidth,trim={50 50 50 50}]{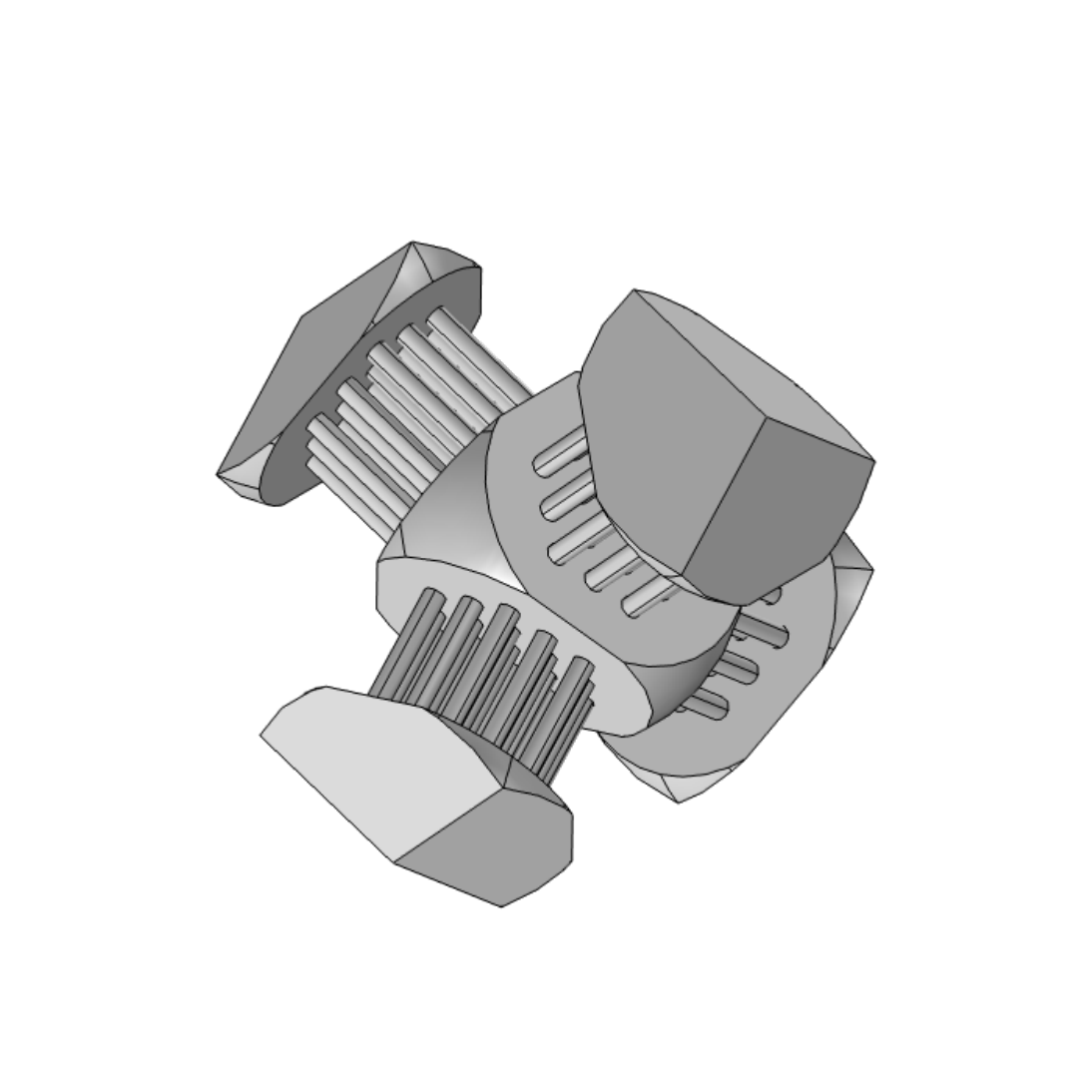}
    \label{fig:mwPrimUnitCell003}}
    \subfloat[][]{
    \includegraphics[height=.5\textwidth, trim={50 50 50 50}]{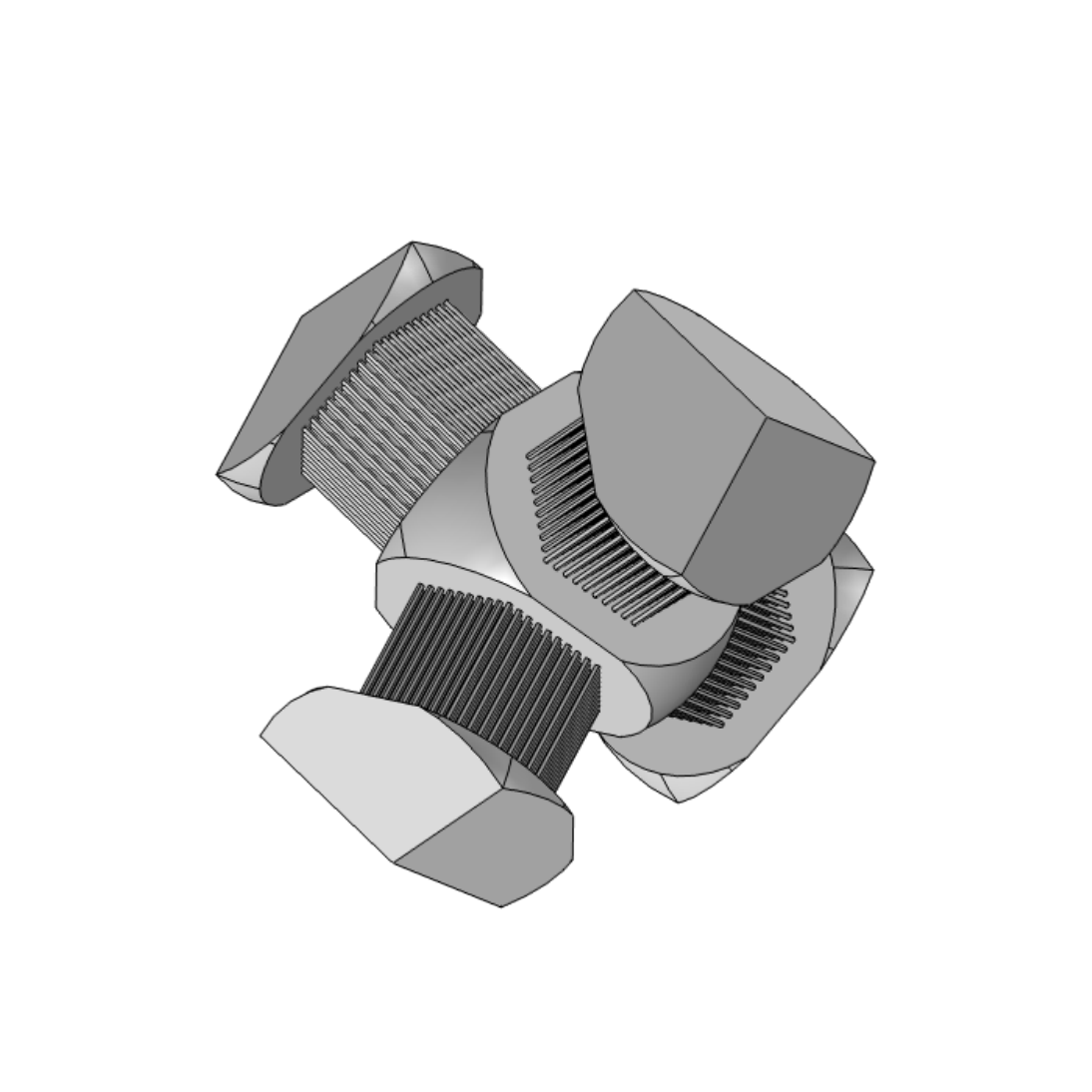}
    \label{fig:mwPrimUnitCell001}}
    \caption{\textit{Metal water} primitive unit cell featured by sheaves of thin beams of diameter (a)~${d_{min} = 3\% \,b}$ and (b)~${d_{min} = 1\% \,b}$.
    \\\seba{e se da qualche parte mettessimo anche la fig con le celle ripetute?}
    }
    \label{fig:mwPrimUnitCell}
\end{figure}

The mutual intersections of the sheaves are accommodated by placing a sphere with a diameter of $ D_{sph} $ at each node of the lattice. The elementary cell consists of four sheaves and two spheres, as depicted in Figure~\ref{fig:mwPrimUnitCell}, with an FCC lattice formed by their spatial repetition. The illustration shows one complete sphere at the intersection point of the four sheaves and one divided into four quarters positioned at the cell extremities, ensuring proper connection with adjacent cells. \\
The spheres serve a dual purpose: they must be large enough to house the sheaves and provide a means to adjust the cell density. Note that increasing $D_{sph}$ reduces the length $L$ of each bundle, consequently affecting the shear-to-axial stiffness ratio of each fibre ($K_{sh}/K_{ax} \propto L^2 $), thus potentially degrading the PM behaviour. Therefore, trimming the spheres with a thickness $t$ using a plane orthogonal to the sheaves can reclaim space, leaving tetrahedron-like shapes in their place.

With respect to the PMs studied in literature, the mass is mainly located in correspondence of the nodes and not at the middle of each link. On the contrary, the flexibility of the PM is guaranteed by the sheaves and not by the small hinges in the nodes. This affects the dynamics of the cell, as discussed in the following.

\subsection{Effective properties}
\label{subsec:mechProperties}

The 3D periodic repetition of a unit cell gives rise to an infinite lattice whose properties are studied by applying the Bloch-Floquet theorem.

In the first place,
% since the size of unit cells is orders of magnitude smaller than the wavelengths of a typical acoustic scattering problem, local inhomogeneities can be averaged and
the resulting lattice is regarded as a homogeneous elastic solid exploiting LFH.
Therefore, its effective mechanical properties are defined through the quasi-static analysis of a primitive cell \cite{hassani1998review707,hassani1998review719} using the \Comsol{} module.\\
% As already mentioned in Section~\ref{sec:intro},
The resulting lattice is characterised by a cubic symmetry, thus the homogenised elasticity tensor $\C^h$ is described by three independent parameters only~\cite{chaves2013notes}: 
\begin{equation}
    \C^h = \begin{bmatrix}
C_{11} & C_{12} & C_{12} & 0 & 0 & 0 \\
C_{12} & C_{11} & C_{12} & 0 & 0 & 0 \\
C_{12} & C_{12} & C_{11} & 0 & 0 & 0 \\
0 & 0 & 0 & C_{44} & 0 & 0 \\
0 & 0 & 0 & 0 & C_{44} & 0 \\
0 & 0 & 0 & 0 & 0 & C_{44} \\
\end{bmatrix}.
\label{eq:cubSymm}
\end{equation}
The eigenvalues of $\C^h$ are $\{C_{11} + 2C_{12},\, C_{11} - C_{12},\, C_{11} - C_{12},\, C_{44},\, C_{44},\, C_{44}\}$ and the ideal \textit{metal water} is achieved when $ C_{11} = C_{12} = B_0 $, $C_{44}=0$, and the effective density $ \rho^h $ is the one of water $ \rho_0 $.
Thus, we set the minimum diameter of each fibre to $ d_{min} = 3\% \,b $ (Figure~\ref{fig:simplified Cell}) and the cell is tuned by optimising the parameters $ A_{tot} $, $ D_{sph} $, and $t$. Assuming that the designed structure is made of Ti 6Al-4V alloy, the achieved properties are:
\begin{gather}    \label{eq:num_stiffness}
    \frac{\C^h}{B_0}=
    \begin{bmatrix}
        \num{0.998} & \num{0.949} & \num{0.949} & \hspace{-1ex}\num{-1.58e-5} & \hspace{-2ex}\num{-1.86e-5} & \hspace{-2ex}\num{-1.68e-5}\\
         & \num{0.998} & \num{0.949} & \hspace{1ex}\num{1.38e-5} & \hspace{-2ex}\num{-2.42e-5} & \hspace{0ex}\num{1.20e-5}\\
         &  & \num{0.998} & \hspace{1ex}\num{1.11e-5} & \hspace{0ex}\num{1.78e-5} & \hspace{-2ex}\num{-1.34e-5}\\
         &  &  & \num{0.0702} & \hspace{-2ex}\num{-1.59e-5} & \hspace{-2ex}\num{-1.44e-5}\\
         & \text{sym} & &  & \num{0.0702} & \hspace{-2ex}\num{-1.50e-5}\\
         &  &  &  & & \num{0.0702}\\
    \end{bmatrix},
    \\ \frac{\rho^h}{\rho_0} = 0.997\text{.}
\end{gather}    
% Furthermore, \textit{metal water}, like water, has to be a PM, thus, it must be able to withstand only one compressional mode with no shear resistance; accordingly, its elasticity tensor $\C$ needs to have rank one, which means that it has only one non-zero eigenvalue \cite{norris2009acoustic}. In order to validate these requirements, Equation \eqref{eq:HookesLawLamé} and Equation \eqref{eq:num_stiffness} are combined to compute $\lambda$ and $\mu$, which are needed to estimate the bulk modulus $B$ and the shear modulus $G$.
We neglect the terms smaller than \num{3e-5} and highlight that:
\begin{align}
    C_{11}=\qty{99.8}{\percent}\, B_0,
    \,
    C_{11}-C_{12}=\qty{4.9}{\percent}\, B_0,
    \,
    C_{44}=\qty{7.0}{\percent}\,B_0,
    \,
    \rho^h=\qty{99.7}{\percent}\,\rho_0;
    \label{eq:properties}
\end{align}
Therefore, the optimised lattice can be referred to as \textit{metal water} in the quasi-static regime. Please note that this approximation improves as the minimum fibre diameter $d_{min}$ decreases. For example, if $d_{min} = \qty{1}{\percent} \,b$
\begin{align}
    C_{11}=\qty{99.9}{\percent}\,B_0,
    \,
    C_{11}-C_{12}=\qty{0.3}{\percent}\,B_0,
    \,
    C_{44}=\qty{0.6}{\percent}\,B_0,
    \,
    \rho^h=\qty{100}{\percent}\,\rho_0.
    \label{eq:properties2}
\end{align}
% It emerges that the modelled \textit{metal water} traces the quasi-static behaviour of $H_2O$ adequately; as a matter of fact, the ratio $B/B_0$ is very close to the unit, the estimated Poisson's ratio $\nu=\frac{\lambda}{3B-\lambda}\approx0.5$, as it happens for fluids, and, finally, the ratio between shear and bulk moduli is very low  $\left(G/B<2\%\right)$, so that a satisfactory pentamode behaviour is achieved.

% Lastly, this optimised primitive unit cell is characterised by a density $\rho=\SI{1000.4}{kg/m^3}$, consenting to achieve a proper matching with the one of pure $H_{2}O$ $\left(\rho_{0}=\SI{1000}{kg/m^3}\right)$. Hence, the designed \textit{metallic water} can be firmly considered a well-functioning water-like PM.

\begin{figure} 
    \centering
    \subfloat[][]{
    \includegraphics[height=.5\textwidth,trim={0 160 0 0}]{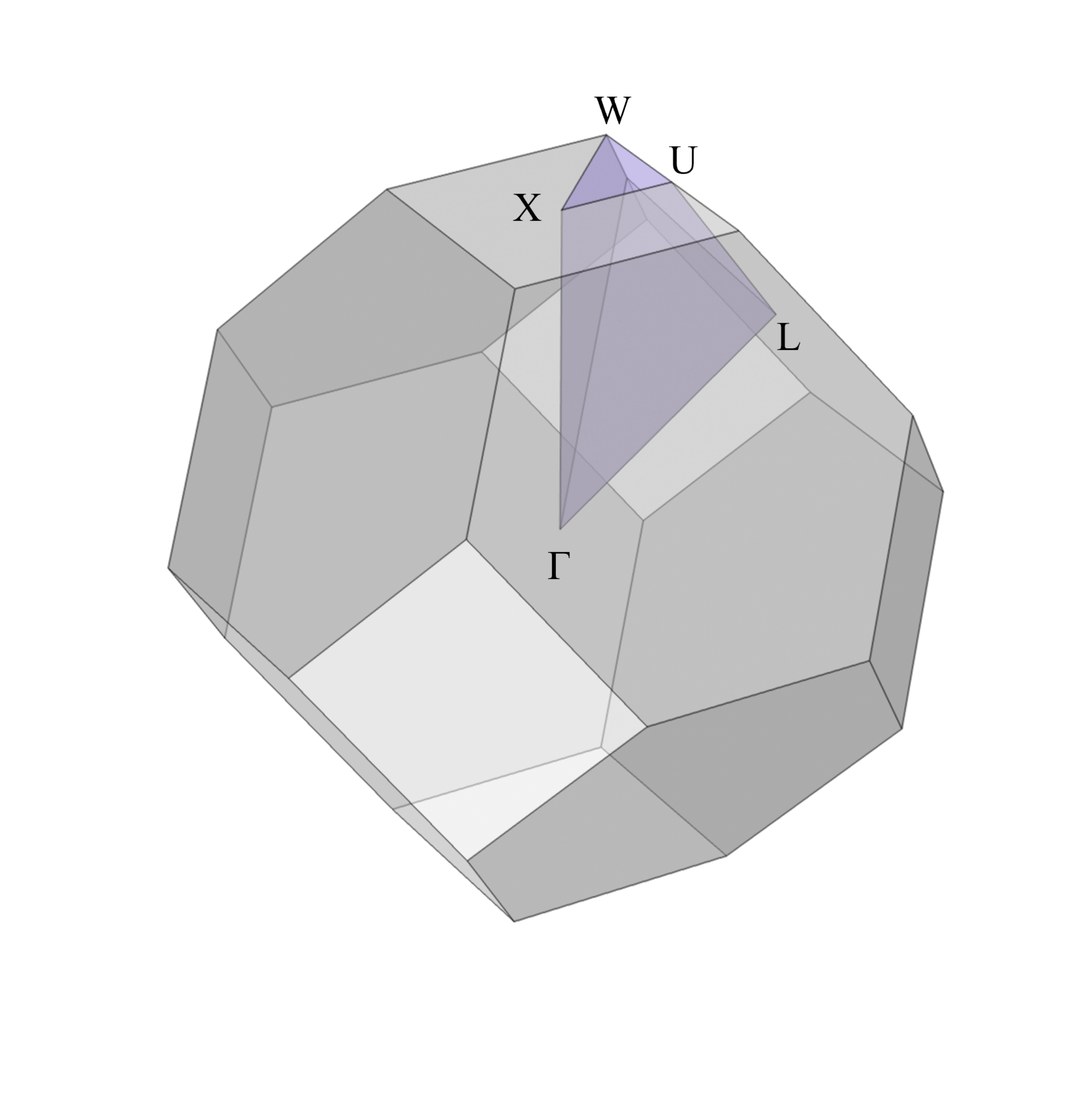}
    \label{fig:firstBZ}}
    \subfloat[][]{
    \includegraphics[height=.5\textwidth,trim={0 0 0 0},clip]{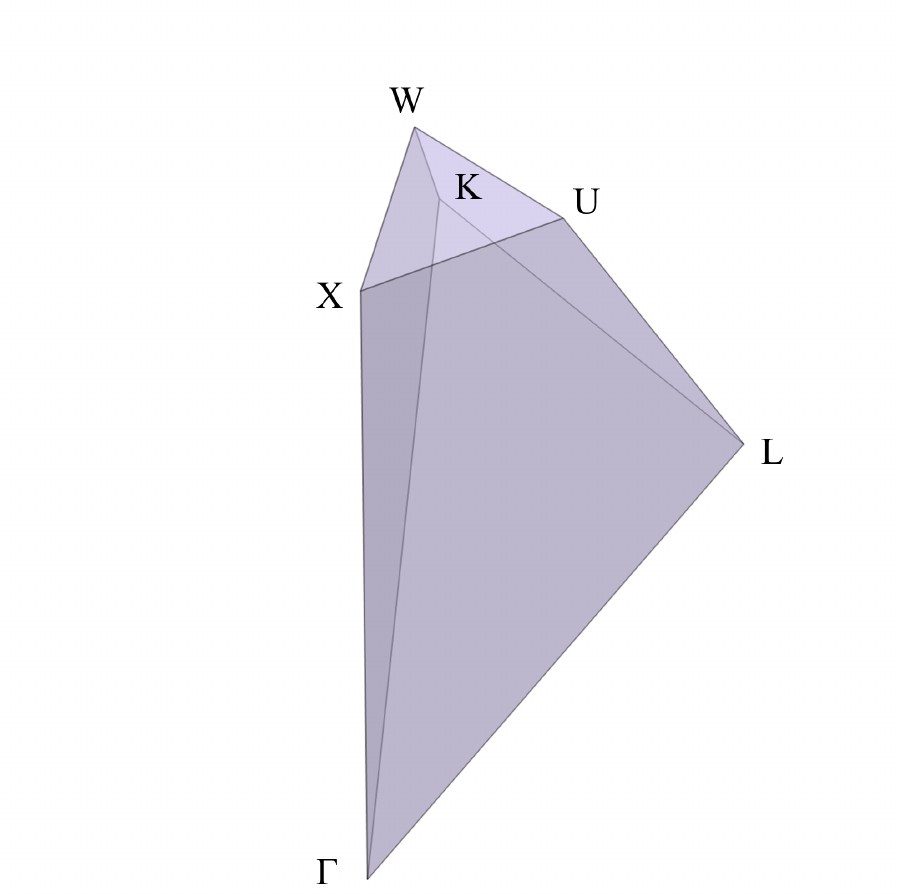}
    \label{fig:IBZ}}
    \caption{(a)~First Brillouin Zone (truncated octahedron) and (b) Irreducible Brillouin Zone of the trigonal-trapezohedral cell.}
    \label{fig:Brillouinzone}
\end{figure}

\begin{figure}
    \centering
    \subfloat[]{
    \includegraphics[height=0.45\textwidth,trim=2 15 280 400,clip]{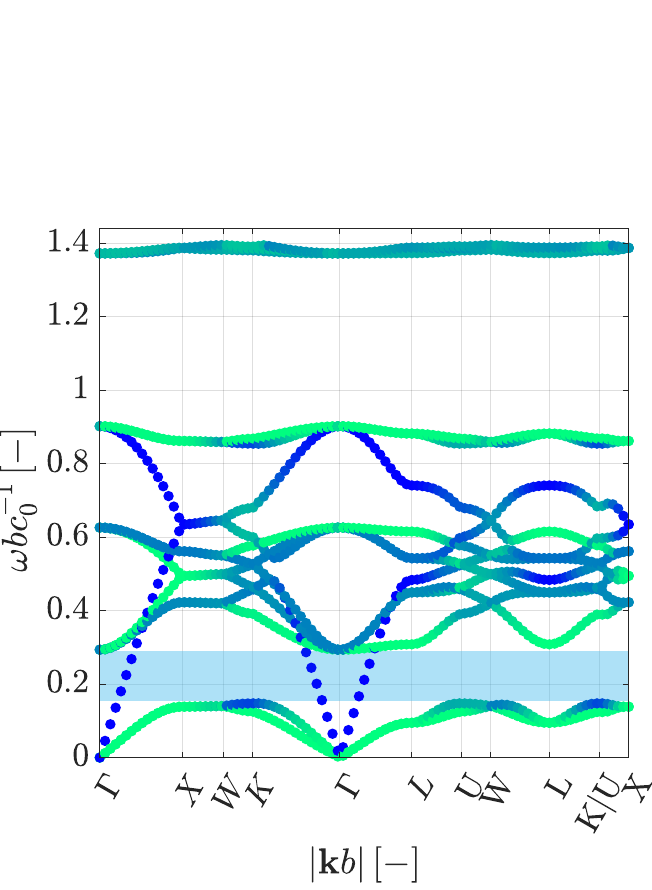}
    \label{fig:dispersionDiagram003_2}}
    \subfloat[]{
    \includegraphics[height=0.45\textwidth,trim=45 15 280 400,clip]{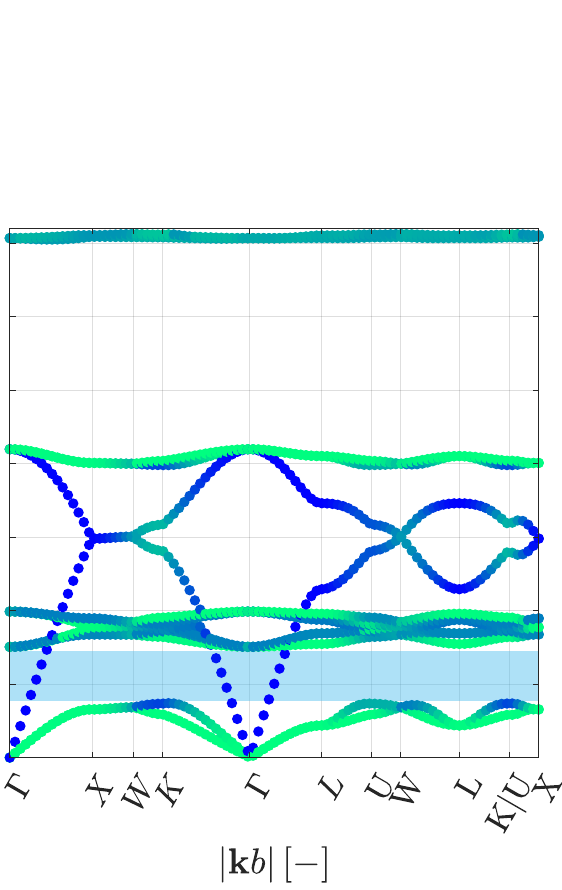}
    \label{fig:dispersionDiagram003_125}}
    \subfloat[]{
    \includegraphics[height=0.45\textwidth,trim=45 15 228 400,clip]{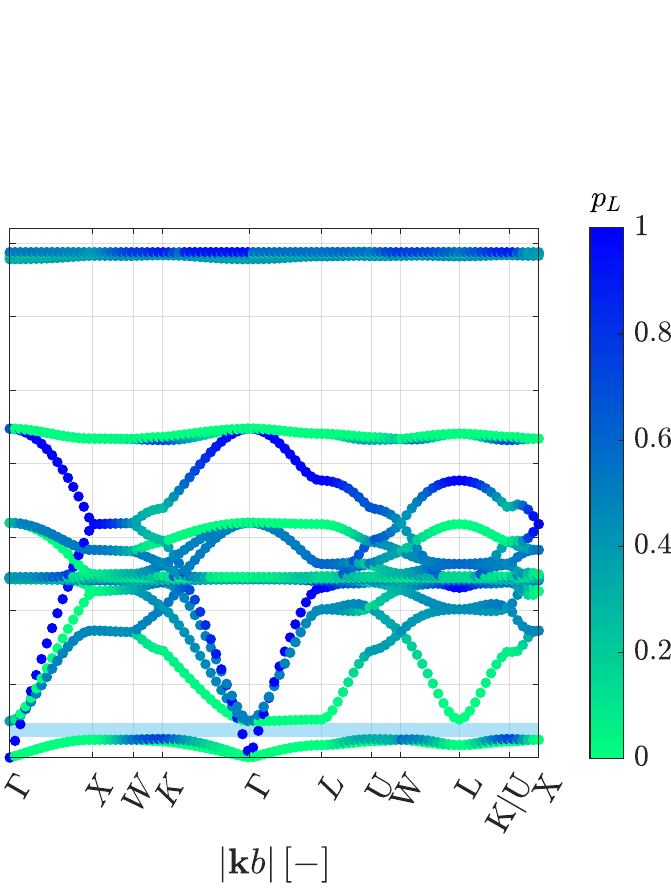}
    \label{fig:dispersionDiagram001_2}}
    \caption{Effect of $d_{min}$ and $l$ on the dispersion diagram of the \textit{metal water} unit cell: (a) $ d_{min} = 3\% \,b$, $l=2d$; (b) $ d_{min} = 3\% \,b$, $l=1.25d$; (c) $ d_{min} = 1\% \,b$, $l=2d$. In (a) and (b) the PM bandgap is at $\omega=\left(0.15\div 0.29\right)c_0b^{-1}$, whereas in (c) it arises at $\omega=\left(0.055\div 0.095\right)c_0b^{-1}$.}
    \label{fig:dispersionDiagram}
\end{figure}

The performance of PMs is fully revealed in their dynamic regime, so the dispersion diagram is computed on the edges of the Irreducible Brillouin zone depicted in Figure~\ref{fig:Brillouinzone}. The polarisation $p_L$ has been introduced as
\begin{equation}
    p_{L}\coloneqq\frac{\int_V{\left|\bb u\cdot\bb k/|\bb k|\,\right|^2\,dV}}{\int_V{|\bb u|^2\,dV}}
\end{equation}
to identify the longitudinal waves, i.e.\ parallel to the wavevector, where $\bb u$ is the complex-valued eigen-displacement and $V$ is the volume of the unit cell.
With reference to Figure~\ref{fig:dispersionDiagram}, the polarisation highlights that only compressional waves propagate (blue branches) within the shear band gap, whereas transverse modes (green branches) experience Bragg’s scattering. Hence, a shear wave band gap opens at $\omega=0.15\,c_0b^{-1}$, due to the low shear stiffness of the lattice.\\
The closure of the band gap occurs at $\omega =0.29\, c_0 b^{-1}$ due to the overlap between the acoustic branch and six branches related to the local rotations of the spheres, reminiscent of the behaviour observed in micropolar materials~\cite{eringen1970foundations}.
On the one hand, the branch starting at $\omega=0.29\,c_0b^{-1}$ is related to the modeshape where the spheres rotate in-phase, so its frequency depends on the shear stiffness of the bundle and decreases whenever tightener bundles are designed, i.e.\ as $l$ is smaller causes smaller $I_\infty$. Refer to Figure~\ref{fig:dispersionDiagram003_2}.\\
On the other hand, the branch starting at $\omega=0.6\,c_0b^{-1}$ occurs when two adjacent spheres counter rotate, this motion is affected by the flexural stiffness of the bundles and decreases when thinner fibres are adopted, Figure~\ref{fig:dispersionDiagram001_2}.\\
A complete band gap opens at higher frequencies due to the local resonance of the spheres in a spring-mass chain fashion.
Lastly, several horizontal branches superimpose on the top of the computed dispersion diagram, due to the resonances localised in the thin fibres composing each sheaf.

\section{Conclusions}
\label{sec:conclusions}

Refining the canonical lattice structures typically employed in PM design, we have developed a 3D metamaterial with a bulk modulus surpassing existing literature values by two orders of magnitude.

Initially, we identify the limitations of conventional 3D PM designs relying on double cone-shaped links, elucidating the challenges in achieving high bulk moduli. 
Thus, we introduce a pioneering approach to PM design, maintaining the lattice configuration while altering the shape of the constituent links. Drawing inspiration from the mechanics of ropes, our novel link shape distributes shear compliance along the links and mass at lattice nodes, diverging from established literature methodologies.
Our innovation culminates in the creation of a PM exhibiting bulk modulus and mass density akin to water. 

By manipulating the geometrical configuration of the artificial unit cell, such a PM can be enriched by anisotropic properties, broadening its applicability to numerous fields requiring analogous properties to liquids.
This milestone represents a paradigm shift, opening avenues for PM utilisation across diverse applications, particularly in acoustics for purposes such as lensing and cloaking.

%\appendix{}
%\section{Appendix}

%\label{Appendix_A}
%\include{Sections/6_A_3D_Sheaf}
%\include{Sections/6_A_Hexagonal}

\bibliographystyle{RS}
\bibliography{main}

\end{document}